\documentclass[
  showpacs,
  twoside,
  twocolumn,
  aps,
  prlsds
]{revtex4}

\usepackage[ngerman,american]{babel}
\usepackage{graphicx}
\usepackage{amsmath}
\usepackage{pxfonts}
\usepackage{wasysym}
\usepackage{manfnt}
\usepackage{color}
\usepackage{dingbat}
\usepackage{multirow}
\usepackage{diagbox}
\usepackage{epstopdf}

\begin{document}

\title{The elastic and directed percolation backbone}
\author{Youjin Deng$^1$}
\email[]{yjdeng@ustc.edu.cn}
\author{Robert M. Ziff$^2$}
\email[]{rziff@umich.edu}
\affiliation{$^1$ Hefei National Laboratory for Physical Sciences at Microscale, Department of Modern Physics, University of Science
and Technology of China, Hefei 230027, China}
\affiliation{$^2$Center for the Study of Complex Systems and Department of Chemical Engineering, University of Michigan, Ann Arbor, Michigan 48109-2136, USA}

\begin{abstract}
We argue that the elastic backbone (EB) (union of shortest paths) on a cylindrical system, recently studied by  Sampaio Filho et al.\ [Phys. Rev. Lett. {\bf 120}, 175701 (2018)], is in fact the backbone of two-dimensional directed percolation (DP).  We simulate the EB on the same system as considered by these authors, and also study the DP backbone directly using an algorithm that allows backbones to be generated in a completely periodic manner.  We find that both the EB in the bulk and the DP backbone have a fractal dimension of $d_{b} = d_{B,\rm DP} = 1.681\,02(15)$ at the identical critical point $p_{c,\rm{DP}} \approx 0.705\,485\,22$.  We also measure the fractal dimension at the edge of the EB system and for the full DP clusters, and find $d_e = d_{\rm DP} = 1.840 \, 54 (4)$.  We  argue that those two fractal dimensions follow from the DP exponents as   $d_{B,\rm DP} = 2-2\beta/\nu_\parallel = 1.681 \, 07 2 (12) $ and $d_{\rm DP} = 2-\beta/\nu_\parallel = 1.840\, 536  (6)$.  Our fractal dimensions differ from the value 1.750(3) found by Sampaio Filho et al.
\end{abstract}



\maketitle


Backbones in percolation systems play a central role in conductivity, permeability, and elasticity, and have been the subject of much research (i.\ e.,  \cite{HerrmannStanley84,Larsson87,Saleur92,DengBloeteNienhuis04,ZhouYangDengZiff12}).  Recently, Sampaio Filho et al.\ \cite{SampaioFilhoEtAl18} have studied the elastic backbone (EB) in percolating systems.  The EB was originally introduced by Herrmann et al.\ \cite{HerrmannHongStanley84} as the shortest path or ensemble of shortest paths, all the same length, between two points.  In \cite{SampaioFilhoEtAl18}, this was  generalized  to a cylindrical system---that is, a square with periodic boundaries in the horizontal direction and open boundaries on the top and the bottom, with the EB defined as the union of shortest paths between occupied sites on the top and occupied sites on the bottom.  Here sites are occupied with probability $p$ as in ordinary percolation and path(s) from the top to the bottom are identified.

Below the ordinary percolation transition point $p_c = 0.592\,746\,05\ldots$ \cite{Jacobsen15,YangZhouLi13,FengDengBlote08}, there are no crossing clusters in the system 
 and therefore no paths.  For $p_c < p < p_{c,\mathrm{EB}}$, the authors of Ref.\  \cite{SampaioFilhoEtAl18} find that there exists of the order of one shortest path.  Above a second transition point at $p = p_{c,\mathrm{EB}} > p_c$, they found the very interesting result that the fraction of the sites in the system that belong to the EB grows rapidly and gives rise to an apparent second-order transition.  For site percolation on a  square lattice rotated by 45$^\circ$, they find $p_{c,\mathrm{EB}} = 0.7055(5).$

In fact, the value  $p_{c,\mathrm{EB}} \approx 0.7055$ coincides with the site directed percolation (DP) threshold, whose value is $p_{c,{\rm DP}} = 0.705 \, 485 \, 22(4)$ \cite{Jensen99,EssamGuttmannDeBell88,WangZhouLiuGaroniDeng13}.  That this process corresponds to DP can be understood by the following simple argument: when the DP threshold is reached, it is possible to have a path that steps down occupied sites at each value of $y$, and therefore has the minimum total length of $L$ on the $L \times L$ lattice.  But once one such backbone can be generated, other equal minimal paths can also occur.  Especially, when $p > p_{c,{\rm DP}}$, there will be multiple backbones, and an apparent transition will be observed. 

\begin{figure}[htbp] 
  \centering
  \includegraphics[width=\columnwidth]{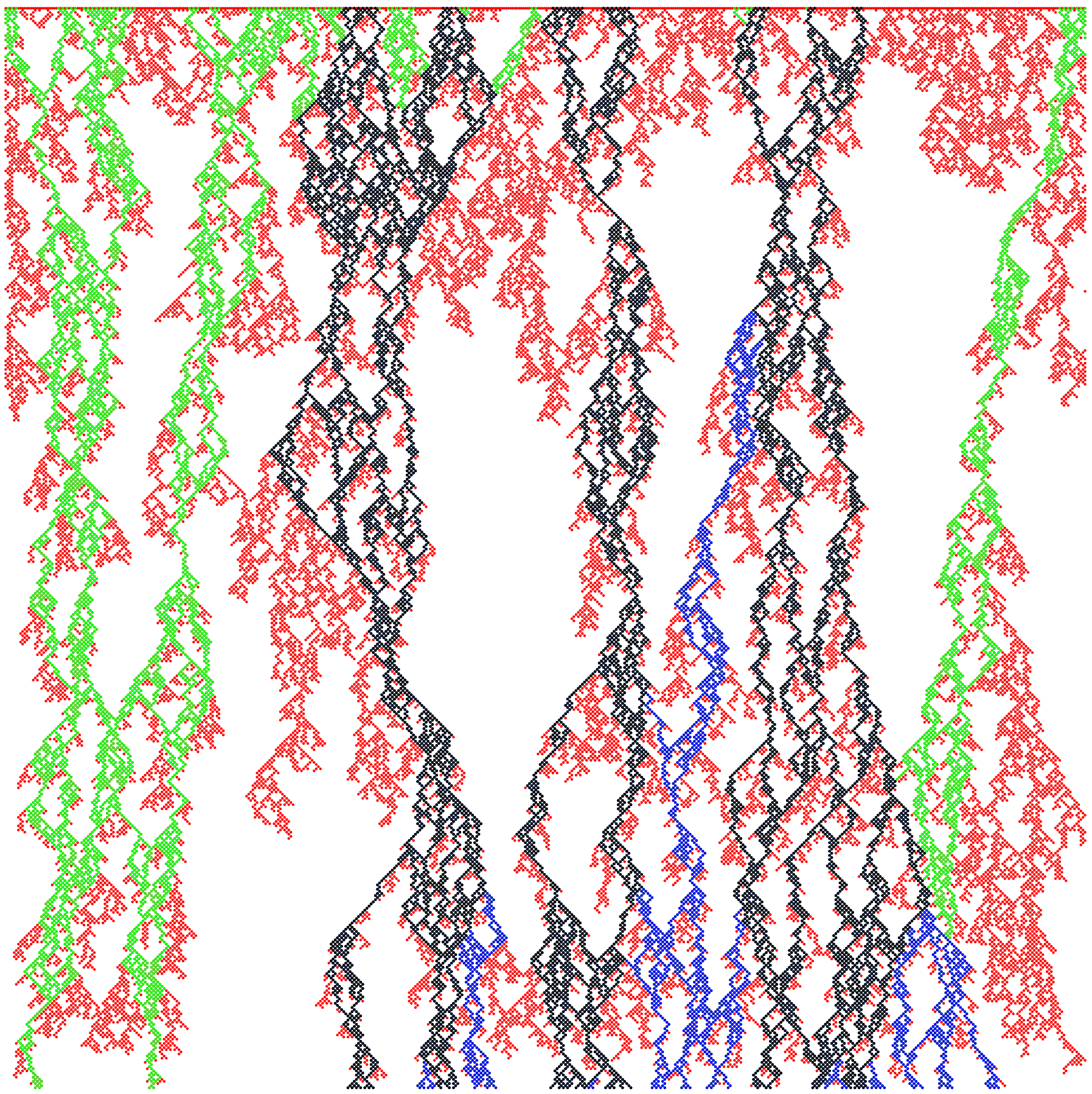} 
  \caption{(color online) Directed percolation (DP) clusters grown from the top row of a 512$\times$512 system (all colors).  Red sites belong to dangling branches that do not reach the bottom.  Blue sites are parts of clusters that do reach the bottom but do not wrap around at the top.  Likewise green sites are parts of clusters that start from the top but do not wrap to sites on the bottom row.  This leaves the black sites which are DP backbones that exist independent of the direction of growth and wrap around periodic boundaries in both vertical directions.}
  \label{fig:dirperc}
\end{figure}

To show that this hypothesis is valid, we study both the original EB model discussed by Simpaio et al., and directly  the backbones of DP, which are shown in Fig.\ \ref{fig:dirperc}.  We find consistent behavior in both of the studies and find values of fractal dimensions that appear to be new, and differ from the fractal dimension found in Ref.\ \cite{SampaioFilhoEtAl18}.




{\bf The elastic backbone}.   As in \cite{SampaioFilhoEtAl18},
we consider site percolation on an $L \times L$ ``tilted" square lattice with open and periodic boundary conditions along the vertical and horizontal directions, respectively.  For this set of simulations we represent the lattice as an $L \times L$ square  of $L^2$ vertices, with diagonals, and the effective aspect ratio (height to circumference) is $1/2$. 
We randomly occupy each lattice site with probability $p$,
construct the cluster from the top edge by the breadth-first search algorithm,
and then ``burn out" those ``dangling" sites which do not belong to any shortest path  connecting the top and the bottom edge.
We measure the following observables as a function of site probability $p$:
\begin{itemize}
\item the spanning probability $P \equiv \langle {\mathcal P} \rangle $ connecting the top and bottom boundaries.
\item the shortest-path length $S \equiv \langle \ell \rangle $ averaged over all realizations including non-spanning ones, where we set $\ell = 0 $ for ${\mathcal P} = 0 $.
\item the average number of occupied EB sites at the top and bottom edges $N_{e} \equiv \langle  ({\mathcal N}_{y=1}+{\mathcal N}_{y=L})/2 \rangle $.
\item the number of occupied EB sites along the centerline, representing the behavior in the bulk $N_{b} \equiv \langle  {\mathcal N}_{y=(L+1)/2}  \rangle$. \item the average number of occupied sites per row in all the rows of the EB, $N_{a}$.
\end{itemize}
   
\noindent
The system undergoes a standard percolation transition at $p_{\rm c}$
where the spanning probability $P$ jumps to a value less than 1 (depending upon geometry), the number of shortest paths connecting the top and bottom edges is ${\mathcal O}(1)$, and their length scales as $S(L) \sim L^{d_{\rm min}}$ with $d_{\rm min} = 1.130\,77(2)$ \cite{ZhouYangDengZiff12,Grassberger92}.

At $p_{c,\rm \mathrm{EB}} \approx 0.7055 $, Sampaio Filho et al.\ observed \cite{SampaioFilhoEtAl18} that the system undergoes another transition  in the dense phase,
at which the number of shortest-paths starts to {\it diverge} and the total mass of the union of the shortest paths scales $N_s \sim L^{d_{\rm s}}$ with a new fractal dimension
$d_{\rm s} = 1.750(3)$.
Together with other observations, the authors claim a novel universality class.

We conjecture that the transition at $p_{c,\rm \mathrm{EB}} \approx 0.7055 $ is {\it precisely equivalent to} site DP on the square lattice,
which has the  threshold at $p_{c, \rm DP}$
and a set of critical exponents as $\beta = 0.276 \, 486(8)$, $\nu_\parallel = 1.733 \, 847(6)$,
$\nu_\perp = 1.096 \, 854(4)$, and $\delta = \beta/\nu_\parallel= 0.159 \, 464(6)$ \cite{Jensen99}.

\vspace{2mm}

\begin{figure}[htbp] 
 \begin{center}
 \includegraphics[width=\columnwidth]{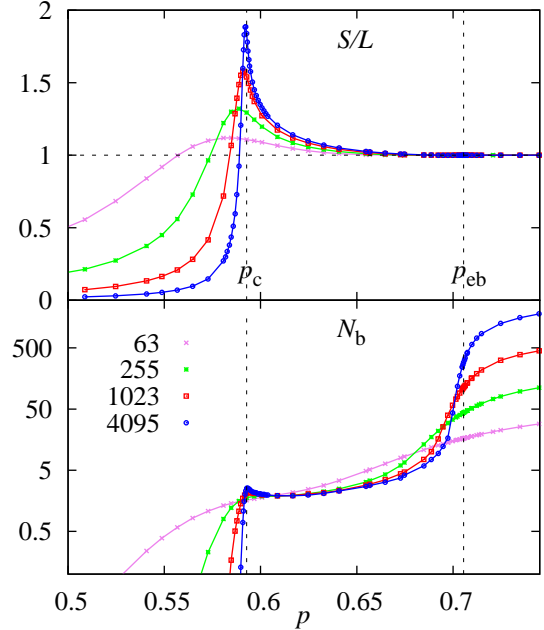}
 \end{center}
 \caption{(color online) The shortest-path length $S$ divided by $L$ (upper plot), and the number of occupied sites along the center row in the bulk $ N_{b}$ as a function of the occupation probability $p$ in the elastic backbone (EB) for the square-lattice
 site percolation (lower plot), for various values of $L$. }
 \label{fig_ElNb}
 \end{figure}

Here we carried out simulations for $L =$ 7, 15, 31,\ldots, 8191 and 15383.  Figure~\ref{fig_ElNb} shows the results for  $S/L$ and  $N_{b}$
as a function of $p$.
It is observed that $S/L$ diverges at the ordinary site threshold $p_{\rm c}$ and converges to a constant for $p > p_{\rm c}$.
Near and above $p_{\rm c, DP}$, $S/L$ is consistent with the value 1 within a statistical uncertainty $ < 10^{-8}$ for $L \geq 63$.
In the whole region $ p_{\rm c} < p < p_{\rm c, DP} $, $N_{b}$ remains constant, suggesting that
the number of paths connecting the top and  bottom boundaries is ${\mathcal O}(1)$. 
For $ p \geq p_{\rm c, DP}$, $N_{b}$ diverges as $L$ increases.
That $S/L$ goes to 1 implies that the backbones simply grow sequentially from $y=L$ to $y=1$ for $ p \geq p_{c, \rm  DP}$.

 \begin{figure}[htbp] 
 \begin{center}
 \includegraphics[width=\columnwidth]{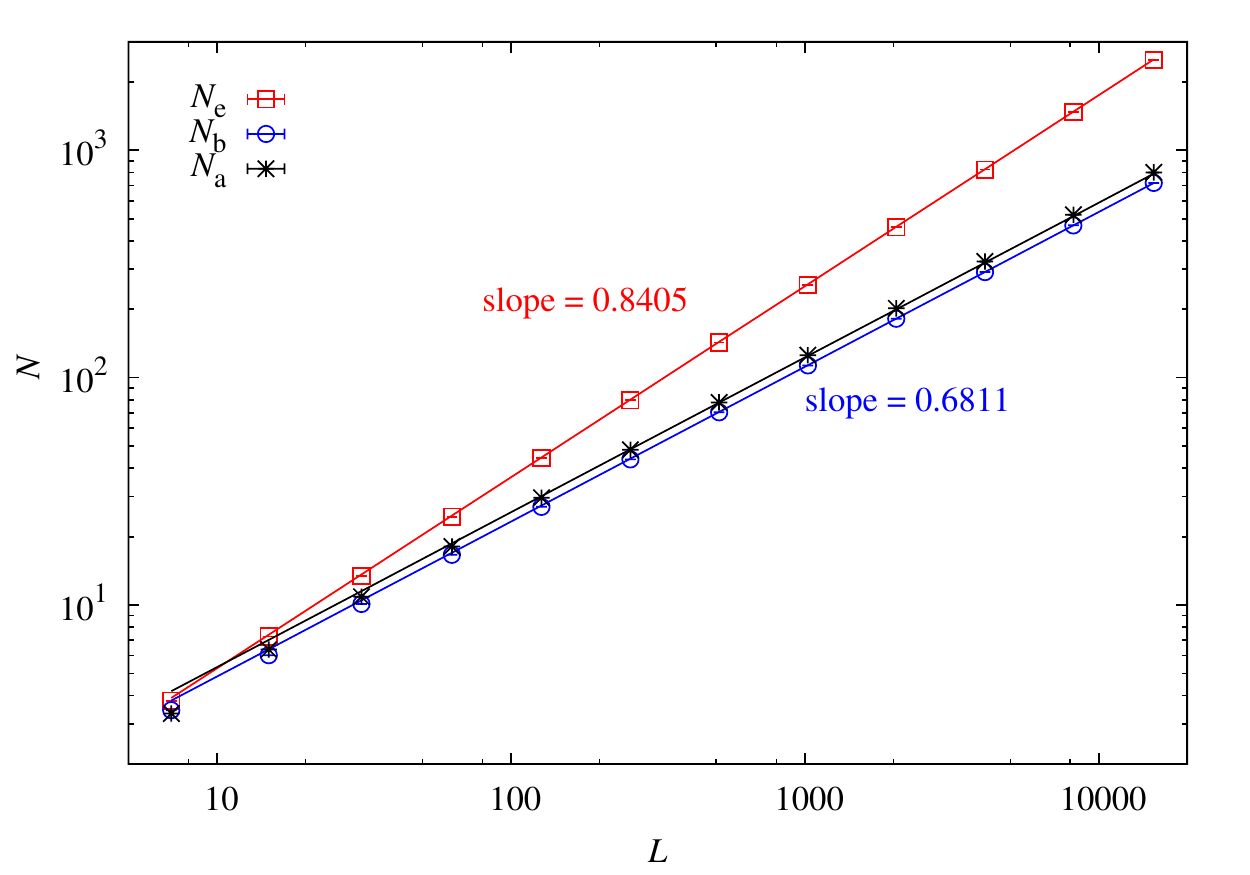}
 \end{center}
 \caption{(color online) $ N_{e}$, $ N_{b}$ and  $N_{a}$ in the EB  at $p_{c, \rm DP} = 0.705 \, 485 \, 22$ {\it vs.}\  $L$, showing the different scaling of the three quantities.  The slopes of the lines are the predicted values from (\ref{eq:DPfractal}) and (\ref{eq:DPbackbonefractal}). }
 \label{fig_NeNbNa}
 \end{figure}

Figure~\ref{fig_NeNbNa} shows  plots of $ N_{e}$, $ N_{b}$ and $N_a$ at $p_{\rm c, DP}$ as a function of $L$,
demonstrating  that their scalings are governed by different exponents.
The least-squares fitting to the ansatz $N_{\mathcal{O}} = L^{d_\mathcal{O}-1} (a+ bL^{-w})$ (see Supplementary Material (SM))
gives $d_{e} = 1.840 \, 54 (10)$, $d_{b} = 1.681 \, 0 2(15)$ $d_{a} = 1.680\,9(2)$.  We subtract 1 from the two-dimensional fractal dimensions because we are taking one-dimensional cuts through the clusters to find $N_e$ and $N_b$. 

\begin{figure}[htbp] 
 \begin{center}
 \includegraphics[width=\columnwidth]{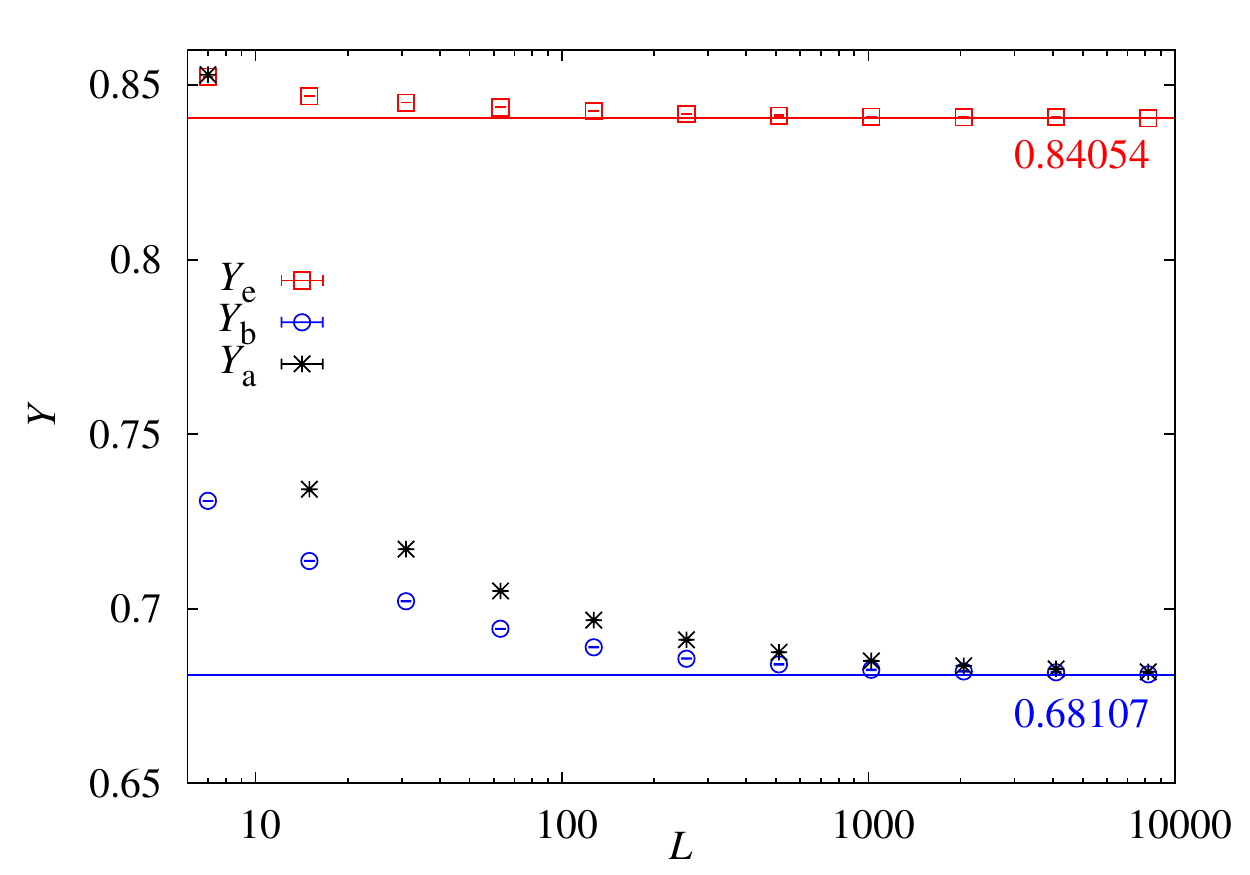}
 \end{center}
 \caption{Effective critical exponents, $ Y_{e} (L)  $, $Y_{b} (L)$ and  $Y_{a} (L)$ for the EB for the square-lattice  site percolation.  These are the local slopes of the data in Fig.\ \ref{fig_NeNbNa}.  The horizontal lines show the predictions of $d_\mathrm{DP} - 1$ from (\ref{eq:DPfractal}) and $d_{B,\mathrm{DP}} - 1$ from (\ref{eq:DPbackbonefractal}). While there are larger finite-size corrections for $Y_a(L)$ than $Y_b(L)$, clearly they both agree with (\ref{eq:DPbackbonefractal}) for large $L$.}
 \label{fig_YeYbYa}
 \end{figure}

We conjecture that the two fractal exponents above can be found from scaling arguments. 
 For the full DP clusters, we expect that  the fractal dimension $d_{\rm DP}$ satisfies the scaling relation \begin{equation}
d_{\rm DP} = 2-\beta/\nu_\parallel = 1.840\, 536  (6)    \label{eq:DPfractal}
\end{equation}
with $\nu_\parallel$ which gives the correlation length in the parallel (time-like) direction.  For the backbones, we have to eliminate the probability of growing finite clusters, so we subtract off the exponent $\delta$ corresponding to the survival probability, $P(t) \sim t^{-\delta}$.  This gives:
\begin{equation}
d_{B,\rm DP}= 2- \beta/\nu_\parallel -\delta = 2- 2\beta/\nu_\parallel= 1.681 \, 07 2 (12)   \label{eq:DPbackbonefractal}
\end{equation}
Indeed, our numerical results for $d_e$ and $d_b$ are in excellent agreement with the DP values above. 
To find the behavior accurately, we considered the local slope or equivalently 
the $L$-dependent exponents as $Y_{\mathcal{O}} (L) = \ln [N_{\mathcal{O}} (L')/N_{\mathcal{O}} (L)]/\ln (L'/L)$
from two consecutive values of $L$. The results, shown in in Fig.~\ref{fig_YeYbYa} in SM, clearly suggest that as $L$ increases,
 $Y_e$ and $Y_b$ converge to the theoretically predicted values.

\begin{figure}[htbp] 
 \begin{center}
 \includegraphics[width=\columnwidth]{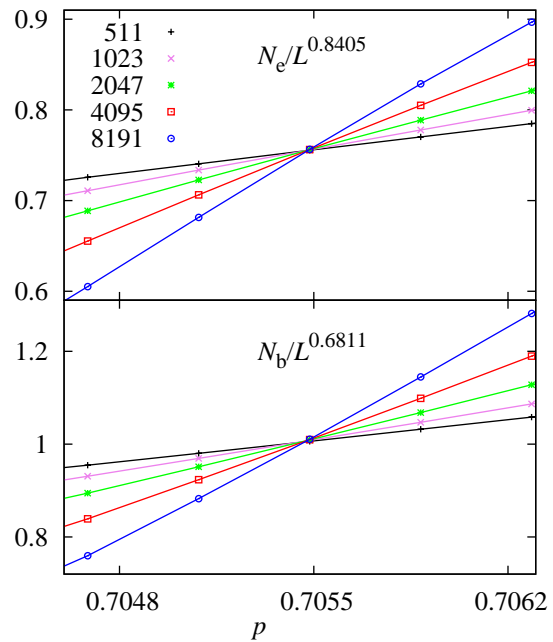}\\
 \end{center}
 \caption{(color online) Number of EB occupied sites at the edge $ N_{e}$ (upper plot) and in the bulk  $ N_{b}$ (lower plot) in the percolation system at $p_{c,\rm DP}$, scaled by exponents $d_{DP} -1 \approx 0.8405$ and  $d_{B,\rm DP}-1 \approx 0.6811$
 from the $(1 \! +\! 1)$-dimensional DP universality class.}
 \label{fig_NeNb}
 \end{figure}
 
Figure~\ref{fig_NeNb} shows the results for $ N_{e}$ and $ N_{b}$ at different $L$,
rescaled by the DP exponents $d_{\rm DP}-1$
and  $d_{B,\rm DP}-1 $,  respectively, as functions of $p$. 
We find nearly perfect crossing at the point $p = p_{\rm c, DP}  = 0.705 \, 485 \, 22$,
giving strong evidence that the transition is at $p_{\rm c, DP}$.
By fixing $d_{\rm DP}$ and $d_{B,\rm DP}$ as in (\ref{eq:DPfractal}) and (\ref{eq:DPbackbonefractal}), the least-squares fitting  gives {$p_c = 0.705 \, 48 5 \, 2 (5)$ and $ \nu_\parallel = 1.72 (2)$} from $ N_{e}$,
and {$p_c = 0.705 \, 485 \, 6 (8)$ and $\nu_\parallel = 1.72 5 (30)$} from $ N_{b}$,  in excellent agreement with (and nearly as precise as) $p_{c, \rm DP}$.

{\bf Directed percolation.}
We also studied DP directly on the cylindrical system, verifying that it has the same properties as the EB at $p_{c,\mathrm{EB}} = p_{c,{\rm DP}}$.  DP has been the object of a great deal of study over the years (i.e., \cite{BroadbentHammersley57,Durrett84,Odor04,WangZhouLiuGaroniDeng13}), although it seems that this aspect of it---the fractal behavior of a collection of multiple clusters and backbones---has not been studied previously.    Backbones of individual DP clusters have been discussed in the context of the de-pinning transition in invasion processes \cite{BuldyrevEtAl92,TangLeschhorn92}.   A single backbone is not a fractal object but an affine one with a roughness exponent of $\zeta = \nu_\perp/\nu_\parallel = 0.6326$.  It is only when one considers the collection of DP backbones that span the system as occurs in the cylindrical system is the fractal nature manifest.  Fractal properties of single DP clusters have been discussed by Kaiser and Turban \cite{KaiserTurban94}. 

To find the DP backbone, one has to trim all branches off that do not make it to the bottom of the system.  Once trimmed of all downward-pointing branches, the DP backbone will be isotropic in both directions,  although it still has a top and bottom edge, similar to what is seen in Fig.\ 1b of Ref.\ \cite{SampaioFilhoEtAl18} for the EB.  In Fig.\ \ref{fig:dirperc} we show the complete clusters of DP (all sites) and the trimmed backbone (all sites except red sites).

We can further make the backbone independent of the horizontal boundaries by continuing to wrap the backbone around, in both upwards and  downwards directions, removing all sites no longer connected.    This procedure yields the black sites in Fig.\ \ref{fig:dirperc}, the blue and green ones being discarded.  This leaves DP backbones that are periodic and having up-down symmetry.   That is, if we started upwards from any row instead of downwards from the top, we would end up with the same EB for the same set of occupied sites.

\begin{figure}[htbp] 
  \centering
  \includegraphics[width=\columnwidth]{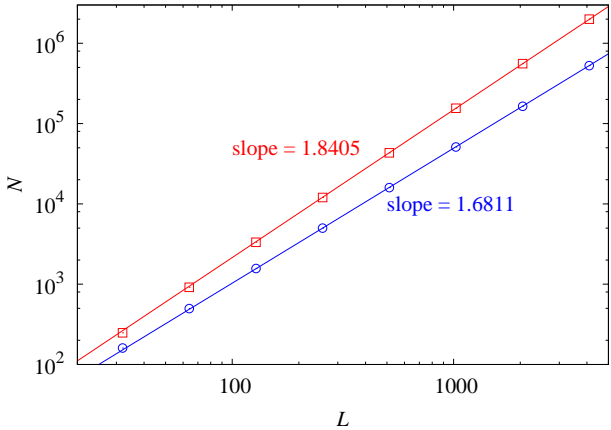}
  \caption{A plot of the average number sites  in DP clusters {\it vs.}\ $L$.  The error bars are smaller than the size of the symbols.  The upper points ($\triangle$) are for $N_{\rm DP}$ of the complete DP clusters (all non-white sites in Fig.\ \ref{fig:dirperc}), while the lower points ($\circ$) are for  $N_{B,\rm DP}$ on periodic DP backbones (black sites in Fig.\ \ref{fig:dirperc}).  Slopes are predicted values from (\ref{eq:DPfractal}) and (\ref{eq:DPbackbonefractal}). Data are given in Table \ref{tab:NDPNB} in SM, and are consistent with these predicted slopes.  }
  \label{fig:DP_NeNb}
\end{figure}

\begin{figure}[htbp] 
  \centering
  \includegraphics[width=\columnwidth]{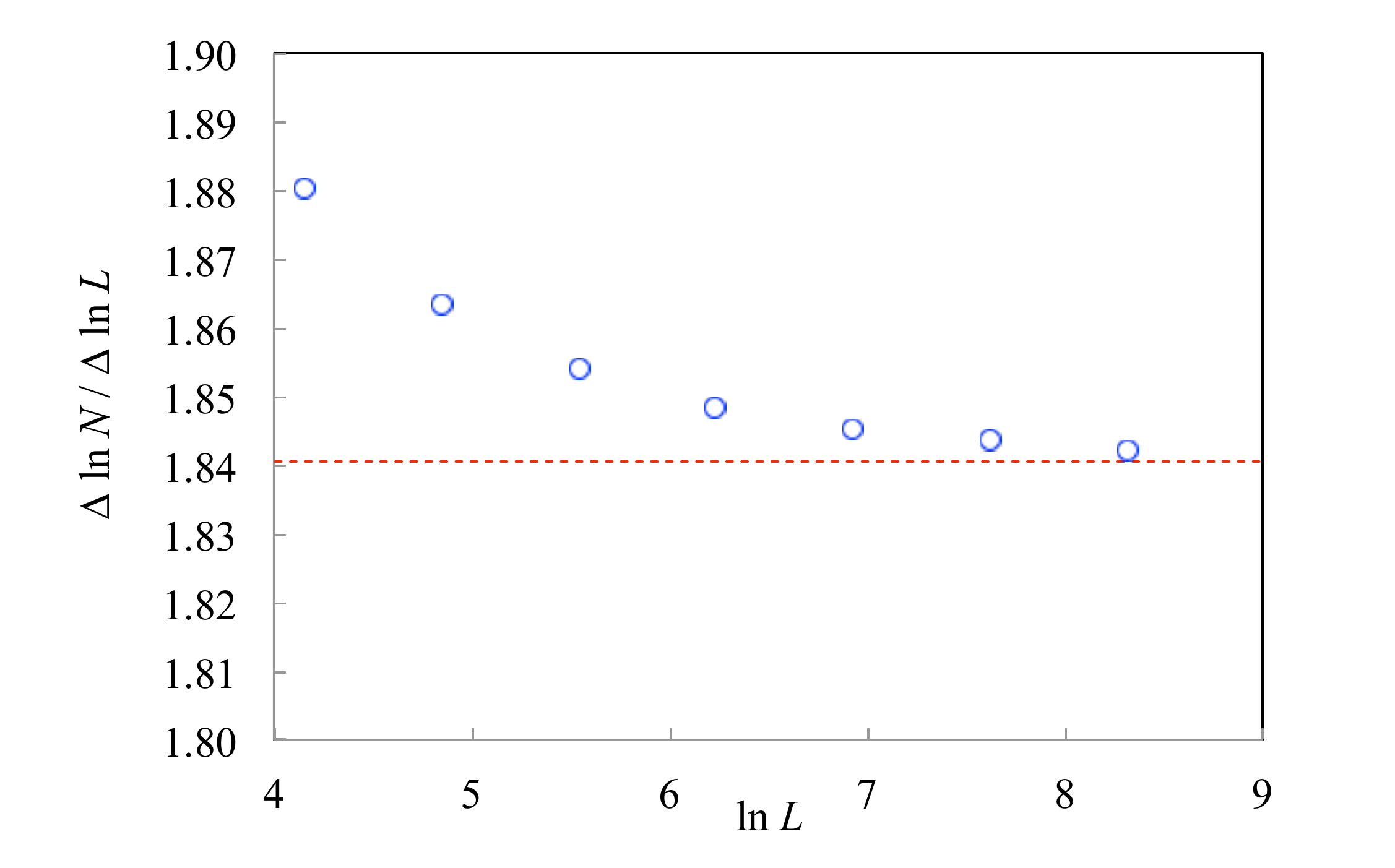}
  \caption{(color online) The local slope of the DP fractal plot of Fig.\ \ref{fig:DP_NeNb}.   The dashed line shows the prediction $d_{\rm DP} = 1.840\,54$ from (\ref{eq:DPfractal}).  }
  \label{fig:DPlocalslope}
\end{figure}

We carried out simulations on $L \times L$ systems with the above trimming procedures.  Here we considered every other site of the square lattice, shifting by one each row, so we considered $L^2/2$ total sites with an aspect ratio of 1. 
We considered $L = 32$, 64,$\ldots,4096$, simulating $1.6 \cdot 10^7$ samples for the five smallest systems, and $5 \cdot 10^6$, $1.5 \cdot 10^6$ and $5 \cdot 10^5$  for the three largest systems.  We measured the number of sites in the full DP clusters, $N_{\rm DP}$,  and the number of sites  in the black backbone $N_{B, \rm DP}$.      These are plotted as a function of $L$ in Fig.\ \ref{fig:DP_NeNb}.  By examining the local slopes, we deduce $d_{b,DP} = 1.680(1)$ (see Fig.\ \ref{fig:DPlocalslope}) in SM and $d_{DP} = 1.842(1)$, and a more careful fitting of the data give $d_{\rm DP} = 1.84066(30)$ and $d_{B,\rm DP} = 1.6815(4)$ (see SM), 
 which precisely agree with the predictions of equations (\ref{eq:DPfractal}) and (\ref{eq:DPbackbonefractal}).

We are not aware of any previous measurement of either of these fractal dimensions.  In comparison, the backbone of ordinary percolation has a dimension of $d_B = 1.643\, 36 (10)$ \cite{XuWangZhouGaroniDeng14}, and the backbone of Eden clusters has a fractal dimension is 4/3 \cite{MannaDhar96}.   The fractal dimension for rigidity percolation is $d_{\rm RP} = 1.78(2)$ \cite{MoukarzelDuxbury95,JacobsThorpe98,MoukarzelDuxbury99}.


{\bf Wrapping probability.}
Crossing and wrapping in DP behaves much differently than in ordinary percolation, where the probability to cross or wrap a system at criticality is less than 1.  (For examples, the probability to cross an open square at criticality is $1/2$ \cite{Cardy92,Ziff92}, to cross an open cylinder is $0.636\, 454$ (aspect ratio 1) or $0.876\,631$ (aspect ratio 1/2) \cite{Cardy06,HoviAharony96,Ziff11}, and the probability to wrap a square torus is $0.521\,058$ \cite{Pinson94,NewmanZiff01}.)  For DP, on the other hand, we find that the probability of crossing or wrapping a cylinder jumps directly to 1 at the critical point.   This is evidently because of  linear behavior of the DP clusters in the large-$L$ limit, which follows from the asymmetry of the exponents $\nu_\parallel$ and $\nu_\perp$.  Clusters are long and thin and thus more easily span the system.

\begin{figure}[htbp] 
 \begin{center}
 \includegraphics[width=\columnwidth]{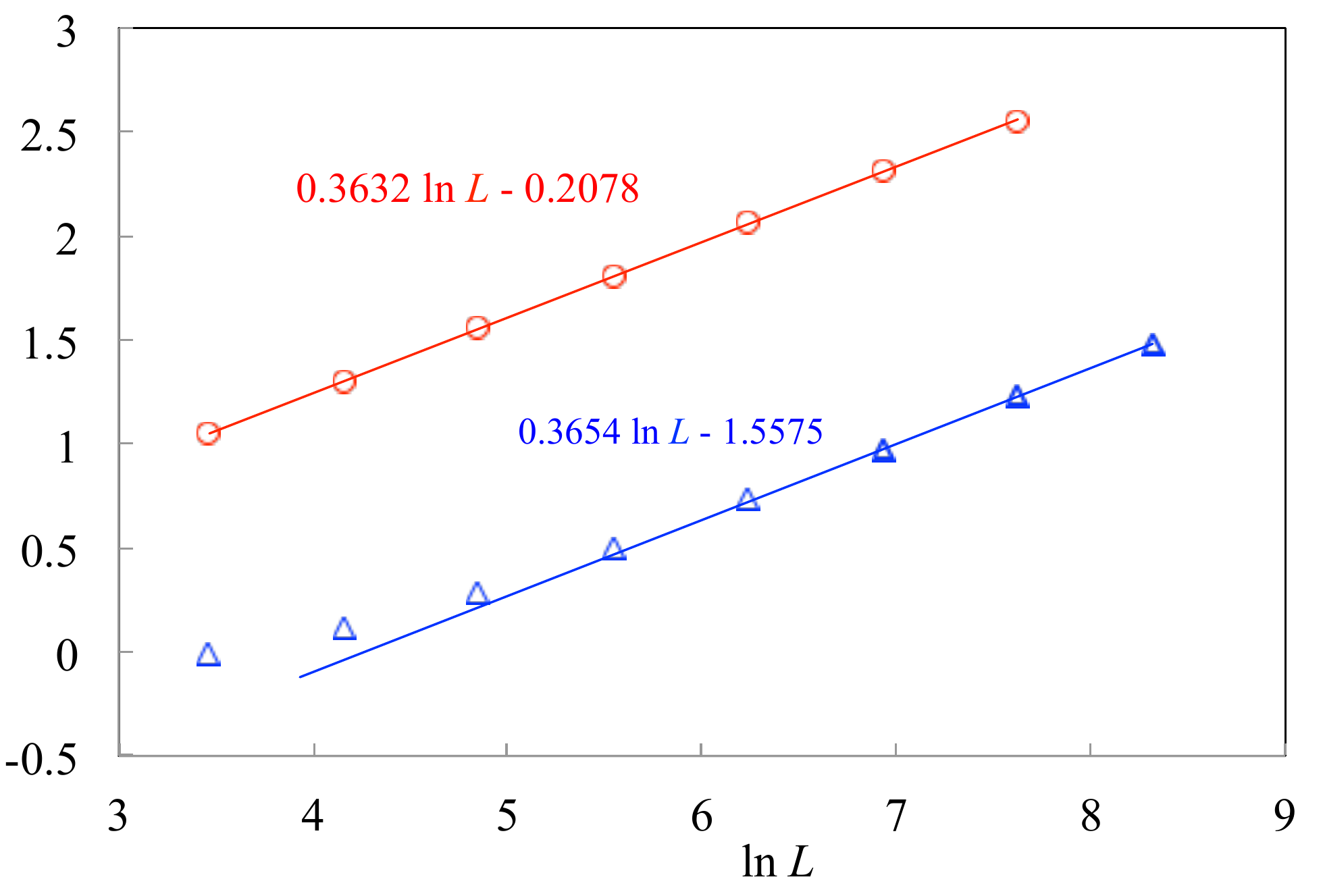}
 \end{center}
 \caption{$\ln[-\ln(1-P_w)]$ {\it vs.}\ $ \ln L$, where $P_w$ is the probability at least one DP backbone wraps the system at criticality ($\circ$), and $\ln \langle N_w \rangle$ {\it vs.}\ $\ln L$, where $\langle N_w \rangle$ is the average number of wrapping DP backbones  ($\triangle$).   Equations give the linear fit through all points (upper plot) and last three points (lower plot).}
 \label{fig:lnln}
 \end{figure}

We measured the probability $P_w$ that at least one DP backbone wraps around the system in the vertical direction in the fully periodic version of our problem (the black backbones in Fig.\ \ref{fig:dirperc}).  We did this for  $L = 32, 64,\ldots,2048$ at $p_{c,{\rm DP}}$, and found values of $P_w$ close to 1, as given in Table \ref{tab:Pw}, SM.  We conjecture that $P_w$ behaves as
\begin{equation}
P_w = 1 - c \exp(-a L^x)
\label{eq:Pw}
\end{equation}
and find that the data are consistent with assuming $c = 1$, which might be expected for a system with periodic b.\ c.\ where there are no surface effects.  Assuming $c = 1$, we can take two logarithms of (\ref{eq:Pw}) to write
\begin{equation}
\ln[-\ln(1-P_w)] = \ln a + x \ln L
\end{equation}
A plot of the data (Fig.\ \ref{fig:lnln}) shows good linear behavior for all $L$ and implies $\ln a = -0.208$ and $x = 0.363$.

It seems reasonable to conjecture that the exponent in (\ref{eq:Pw}) is proportional to the average number of wraparound backbone clusters  $\langle N_w \rangle$ in the system.  Indeed, we measured $\langle N_w \rangle$  as a function of $L$, and plotting on a log-log plot (Fig.\ \ref{fig:lnln}) we see that for large $L$,  $\langle N_w \rangle \sim  e^{-1.558}L^{0.365} = A L^{0.364}$, whose exponent is nearly identical to $x$ in (\ref{eq:Pw}).  This confirms our conjecture about the behavior of  $\langle P_w \rangle$ being related to $\langle N_w \rangle$. 

We also predict that $x = 1-\zeta = 1-\nu_\perp/\nu_\parallel \approx 0.367\, 387$, since the width of clusters grows as $L^\zeta$ and the width of the system grows as $L$, so $\langle N_w \rangle \sim L^{1-\zeta}$.  Our data for both $P_w$ and $\langle N_w \rangle$ (large $L$) are consistent with this value of $x$.  We can put all these results together to write

\begin{equation}
P_w = 1 - q^{\langle N_w \rangle}
\end{equation}
where  $q = \exp(-a/A) = 0.0211 $.  One can interpret $q$ as the probability that a given wraparound backbone cluster does not survive.

\vspace{2mm}
\noindent
{\bf Conclusions.} The authors of Ref.\ \cite{SampaioFilhoEtAl18} have uncovered that buried within supercritical percolation there exists an interesting EB transition which is related to a problem of DP.   The fractal properties of the EB are exactly related to DP critical exponents, which we verify with careful simulations on both the EB and DP directly. The fractal dimension  $d_{S} = 1.750(3)$ found in  \cite{SampaioFilhoEtAl18}  is about halfway between the edge and bulk fractal dimensions (\ref{eq:DPfractal}) and (\ref{eq:DPbackbonefractal}), and this value evidently represents a different quantity than we are measuring here.  This question deserves further investigation.

 {\bf Acknowledgments.} This work was partly supported by the National Science Fund for Distinguished Young Scholars (NSFDYS) under Grant No.\ 11625522 and  the Ministry of Science and Technology of China Grant No.\ 2016YFA0301604 (YD).  The authors thank Hans Herrmann for useful comments on the manuscript.
 
\bibliography{bibliography.bib}

\hrule


\begin{center}

{\bf Supplementary Data}

{\bf The elastic and directed percolation backbone}

Youjin Deng and Robert M. Ziff

\end{center}

\vskip 0.2 in

The Supplementary Material contains additional plots of the data, and tables of the data plotted in some of the figures.

\section{Additional plots}

\begin{figure}[htbp] 
  \centering
  \includegraphics[width=\columnwidth]{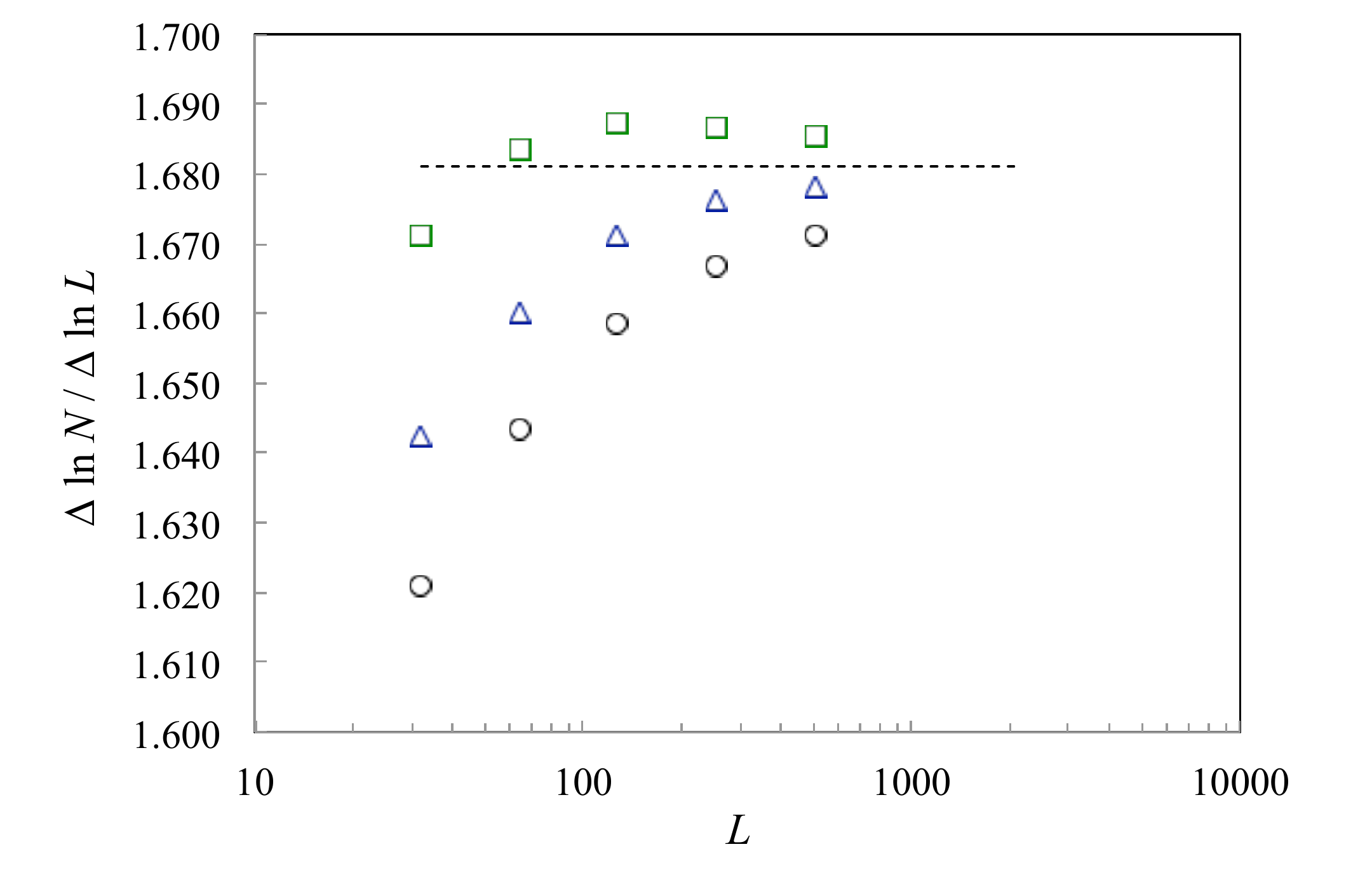}
  \caption{(color online) The local slope of the DP backbone including the blue, green, and black sites ($\square$), green and black sites ($\triangle$), and the just black sites of Fig.\ \ref{fig:dirperc} ($\circ$).   The dashed line shows the prediction $d_{B,\rm DP} = 1.681\,07$ from (\ref{eq:DPbackbonefractal}).  For the black sites, this quantity represents the local slope of the plot in Fig.\ \ref{fig:DP_NeNb}.  It can be seen that all estimates converge to a similar value for large $L$.}
  \label{fig:DPbackbonelocalslope}
\end{figure}

\section{Data}

The following tables show the data given in various plots.

\begin{table}[htbp]
 \begin{center}
   \begin{tabular}{|c|c|c|c|c|c|}
   \hline
      \diagbox{$L$}{$\epsilon$}&$-0.0008$&$-0.0004$&0&0.0004&0.0008\\ \hline
    {511} &137.137(21)&139.948(21)&142.762(93)&145.551(21)&148.340(20)\\
     \hline
     {1023} &240.778(63)&248.495(63)&255.995(21)&263.470(61)&270.818(60)\\
     \hline
    {2047} &417.83(15)&438.70(15)&458.75(4)&478.55(15)&498.13(14)\\
     \hline
    {4095} &712.20(29)&767.54(30)&821.81(6)&875.09(28)&926.26(26)\\
     \hline
     {8191} &1177.52(47)&1326.25(47)&1472.03(11)&1612.47(44)&1745.23(41)\\
     \hline
     \end{tabular}
     \caption{Monte Carlo simulational data for $N_{e}$ plotted in Fig.~\ref{fig_NeNb} for different values of $p = p_{c,{\rm DP}} + \epsilon$ where $p_{c,{\rm DP}} = 0.705\,485\,22$. The numbers in parentheses are error bars in the last digit(s).}
     \label {tab:Ne}
  \end{center}
  \end{table}

 \begin{table}[htbp]
 \begin{center}
   \begin{tabular}{|c|c|c|c|c|c|}
   \hline
   \diagbox{$L$}{$\epsilon$}&$-0.0008$&$-0.0004$&0&0.0004&0.0008\\
    \hline
    {511} &66.774(16)&68.574(16)&70.358(7)&72.168(16)&74.024(16)\\
     \hline
    {1023} &104.469(41)&108.847(42)&113.137(15)&117.496(43)&121.932(43)\\
     \hline
     {2047} &160.972(89)&171.378(92)&181.626(24)&192.316(96)&203.004(95)\\
     \hline
     {4095} &242.13(15)&266.47(16)&291.43(3)&316.97(17)&343.41(17)\\
     \hline
     {8191} &351.57(21)&408.55(22)&467.58(6)&529.76(24)&593.10(25)\\
     \hline
     \end{tabular}
     \caption{Monte Carlo simulational data for $N_{b}$ plotted in Fig.~\ref{fig_NeNb} for different values of $p = p_{c,{\rm DP}} + \epsilon$ where $ p_{c,{\rm DP}} = 0.705\,485\,22$. The numbers in parentheses are error bars in the last digit(s).}
     \label {tab:Nb}
  \end{center}
  \end{table}

\begin{table}[htbp]
\begin{center}
 \begin{tabular}{|c |l |l |l |}
 \hline
  $L$   & $N_{\rm e}$            & $N_{\rm b}$                 & $N_{\rm a}$         \\
 \hline
  7       & $3.795 \, 77(3)  $   & $3.445\, 03(3) $            & $3.328 \, 84 (3) $ \\
 \hline
  15     & $7.268\, 45(7) $     & $6.013\, 07(7)$             & $6.377 \, 14(6) $ \\    
 \hline
  31     & $13.439\, 9(2) $     & $10.095\, 5(2) $            & $10.866 \, 9(2) $  \\
 \hline 
  63     & $24.470\, 2(4) $     & $16.610\, 7(3) $            & $18.069 \, 9(3) $  \\
 \hline
 127    & $44.206\, 2(5) $     & $27.025\, 2 (5)$            & $29.623 \, 4(4) $ \\
 \hline
 255    & $79.529\, 8(14) $   & $43.687\, 1(12)$           & $48.146 \, 1(9) $ \\ 
 \hline
 511    & $142.767(3) $        & $70.365\, 2  (24)$          & $77.840\, 0(18)$  \\
 \hline
1023   & $255.995(7) $        & $113.127(5)$                 & $125.448 (4)$ \\
 \hline
2047   & $458.730(16)$       & $181.620(10)$               & $201.759 (8) $  \\
 \hline
4095   & $821.767(30)$       & $291.451(20)$               & $324.121(12) $ \\
 \hline
8191   & $1472.02(5)$         & $467.57(3)$                   & $520.33(2) $ \\
 \hline
15383  & $2500.12(8) $        & $718.30(4)$                   & $799.69(3) $ \\
 \hline
  \end{tabular}
  \caption{Monte Carlo simulation data for $N_{\rm e}$, $N_{\rm b}$ and $N_{\rm a}$ at $p_{\rm c, DP} = 0.705 \, 485 \, 22$, plotted in Fig.~\ref{fig_NeNbNa}.
  The numbers in parentheses are error bars in the last digit(s).}
  \label {tab:NeNbNa}
\end{center}
\end{table}

 \begin{table}[htbp]
 \begin{center}
  \begin{tabular}{|l|l|}
  \hline
   $L$&$P_{w}$\\
   \hline
   32& $ 0.953\,767\,72 $  \\
  \hline
   64& $0.979\,691\,45 $\\
  \hline
  128& $0.992\,960\,27$ \\
  \hline
  256& $0.998\,210\,31$\\
  \hline
  512& $0.999\,694\,41$ \\
  \hline
 1024& $0.999\,960\,33$ \\
  \hline
 2048& $0.999\,997\,29$ \\
  \hline
   \end{tabular}
   \caption{Monte Carlo simulational data for $P_{w}$, plotted in Fig.~\ref{fig:lnln}. Note: Some of this data will be updated in the final version of the paper.}
   \label {tab:Pw}
 \end{center}
 \end{table}

 \begin{table}[htbp]
 \begin{center}
  \begin{tabular}{|l|l|l|}
  \hline
   $L$&$N_{\mathrm {DP}}$&$N_{B,\mathrm {DP}}$\\
   \hline
   32& $ 248.080$ & $ 170.470 $\\
  \hline
   64 & $ 913.215 $ & $524.268 $\\
  \hline
  128& $ 3322.58$& $1637.62$ \\
  \hline
  256 & $ 12\,010.0$ & $5169.45$\\
  \hline
  512& $ 43\,248.4 $ & $16\,413.1$\\
  \hline
 1024 & $ 155\,396. $& $52\,274.9^*$\\
  \hline
 2048 & $ 557\,679. $ & $164\,291.^*$\\
 \hline
 4096 & $ 1999\,477. $ & $526\,812^*.$\\
  \hline
   \end{tabular}
   \caption{Monte Carlo simulational data for the average number of sites in the DP clusters and in the DP backbones as functions of system size $L$, plotted in Fig.\ \ref{fig:DP_NeNb}.  Errors are expected to be of the relative order of about $10^{-3}$ based upon the number of samples simulated. ($^*$ Some of this data will be updated in the final version of the paper.)}
   \label {tab:NDPNB}
 \end{center}
 \end{table}
 
 \section{Fit of EB and DP data}
 
 For the all the following quantities, we use the anzatz $N=L^{d_f-1}(a+bL^{-w})$ to fit the data:
 
A fit of the EB data for $N_e$ (Table \ref{tab:NeNbNa}) with $L\geq 127$ gives 
$d_e=1.840 54(10)$, $a=0.7562 (7)$, $b=-0.08(2)$, and $w=-0.7(1) $.

A fit of the EB data for $N_b$ (Table \ref{tab:NeNbNa}) with $L\geq 127$ gives 
 $d_b=1.681 02(15)$, $a=1.012(15)$, $b=-0.49(6)$, and $w=-0.73(5)$.
 
A fit of the EB data for $N_a$ (Table \ref{tab:NeNbNa})  with $L\geq 255$ gives $d_a=1.6809(2)$, $a=1.129(3)$, $b=-0.71(10)$, and $w=0.62(4)$. 

A fit of the DP data  (Table \ref{tab:NDPNB})  for all $L$ yields $d_{\mathrm DP}=1.84066(30)$, $a=0.4492(12)$, $b=-0.40(2)$, and $w=0.76(2)$. 

A fit of the DP backbone data  (Table \ref{tab:NDPNB}) for $L \geq 128$, yields $d_{B,\mathrm DP}=1.6817(6)$, $a=0.4434(20)$, $b=5.3(20)$, and $w=1.5(2)$.  

\section{Ratio of number of sites in DP and DP backbone}

\begin{figure}[htbp] 
  \centering
  \includegraphics[width=\columnwidth]{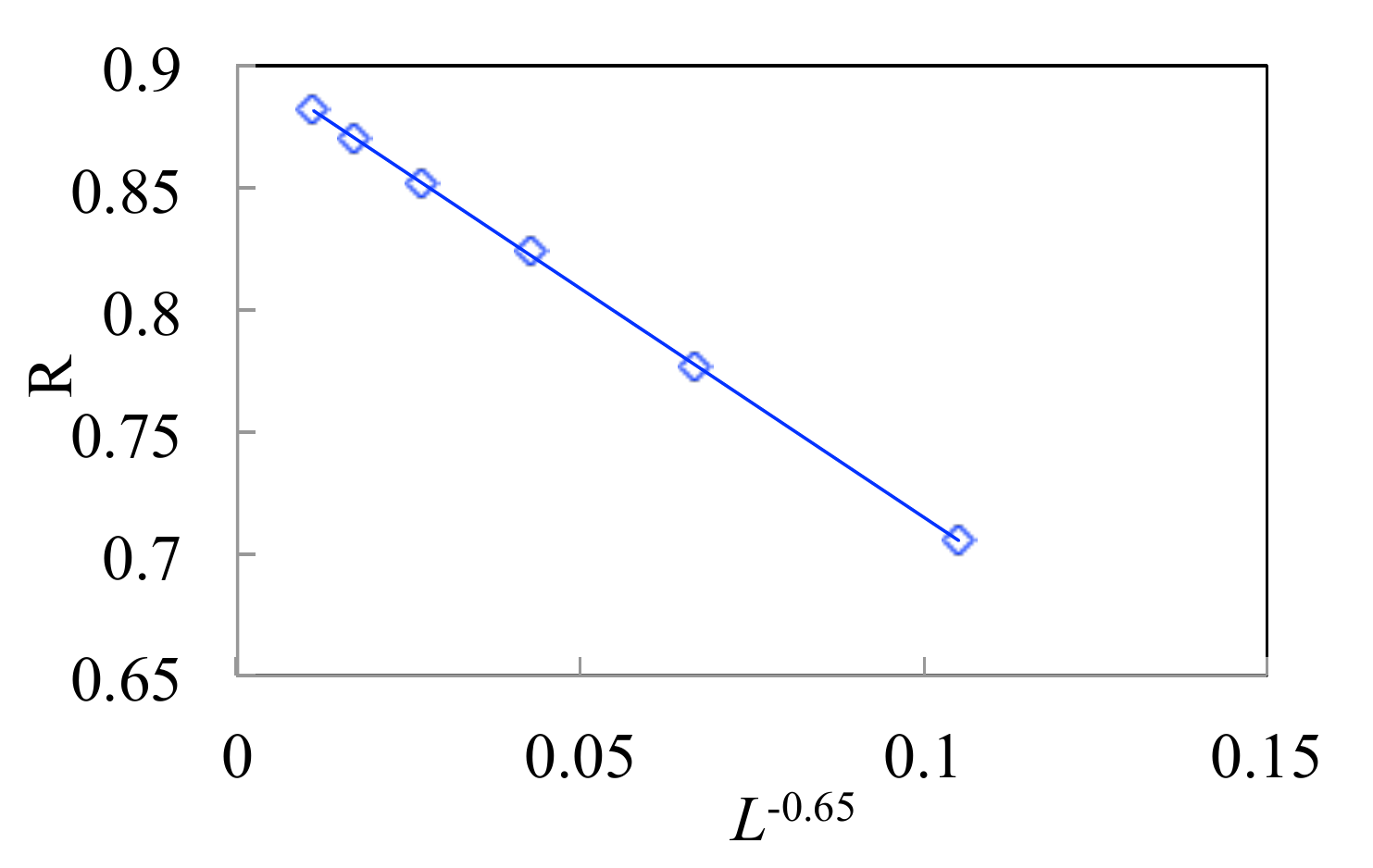}
  \caption{(color online) The DP ratio $R$ of equation (\ref{eq:R}) vs.\ $L^{-0.65}$.  The line represents a linear fit to the data with $R =  0.9023 - 1.874  L^{-0.65}$.}
  \label{fig:ratio}
\end{figure}

Here we add an observation about the scaling of the  numbers $N_{B,\mathrm{DP}}$ and $N_\mathrm{DP}$.
From (\ref{eq:DPfractal}) and (\ref{eq:DPbackbonefractal}), we have the relation
\begin{equation}
d_{B,\rm DP} = 2 d_{\rm DP} - 2 
\label{eq:dB}
\end{equation}
and as a consequence we should expect that the quantity
\begin{equation}
R = \frac{N_\mathrm{DP}^2}{N_s N_{B,\mathrm{DP}}}
\label{eq:R}
\end{equation}
converges to a constant for large $L$, where $N_s$ is the number of sites in the system, which is $L^2/2$ for the geometry used here.  Plotting this $R$ vs.\ $L^{-w}$, we see in Fig.\ \ref{fig:ratio} that we get a good straight line for $w = 0.65$, and that $R$ reaches a constant $\approx 0.9023$ as $L \to \infty$.   (Note here we used the periodic backbone---the black sites---for $N_{B,\mathrm{DP}}$.)  That $R$ goes to a constant provides further confirmation that the scaling relation (\ref{eq:dB}) is valid.    



\bigskip

\section{Fluctuations}

\begin{figure}[htbp] 
  \centering
  \includegraphics[width=\columnwidth]{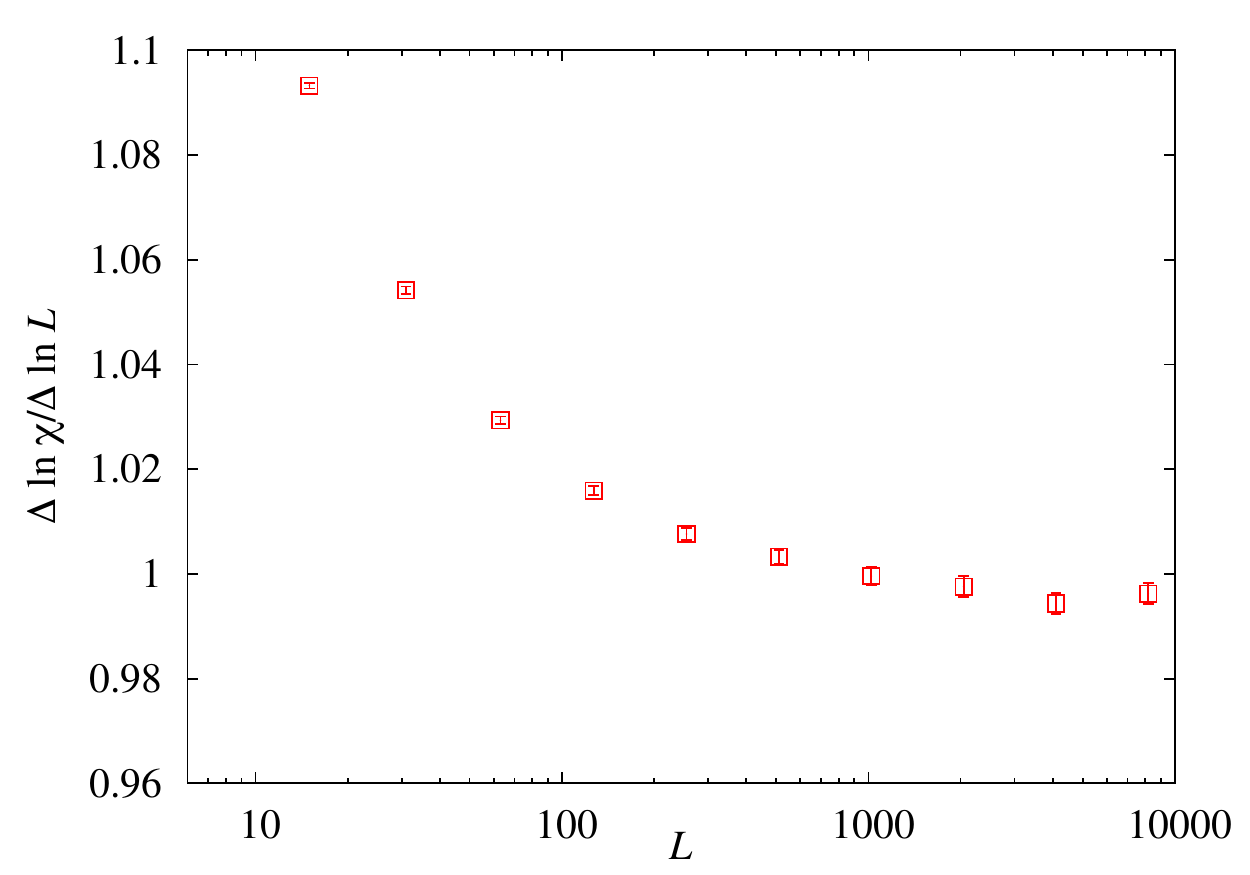}
  \caption{(color online)  Estimates for the scaling dimension of the fluctuations $\chi = \langle N_a^2 \rangle - \langle N_a \rangle ^2$ in the EB.}
  \label{fig:Ychi}
\end{figure}

We also measured the fluctuations in the total number of sites $N_t$ in the EB, $ \chi := L^{-2} ( \langle N_t^2\rangle - \langle N_t\rangle^2) = \langle N_a^2\rangle - \langle N_a\rangle^2.$ The effective scaling dimension appears to be about  0.995(10) as shown in Fig.\ \ref{fig:Ychi}, and is consistent with the value 1.00(2) given in \cite{SampaioFilhoEtAl18}. 

\end{document}